\documentclass[]{aa}
\usepackage{epsfig}
\begin{document}

\newcommand{\nc}{\newcommand}
\nc{\Porb}{$P_{\rm orb}$\,}

\title{On the link between rotation, chromospheric activity and
       Li abundance in subgiant stars
                            }

 \author{   
            J. D. do Nascimento Jr. \inst{1},
            B. L. Canto Martins \inst{1},
	    C. H. F. Melo \inst{2,1},
            G. Porto de Mello \inst{3}
      \and  J. R. De Medeiros \inst{1}
  }
  \institute{
    Departamento de F\'{\i}sica,
    Universidade Federal do Rio
    Grande do Norte, 59072-970
    Natal, RN., Brazil
        \and
    European Southern Observatory, Casilla 19001, Santiago 19, Chile
        \and
    Observat\'orio do Valongo, Ladeira do Pedro Antonio, 43, 20080-090, Rio de Janeiro, RJ., Brazil}

  \offprints{J.D. do Nascimento Jr }
  \mail{do.nascimento@dfte.ufrn.br}

  \date{Received / Accepted}

\abstract{ The connection rotation--CaII emission flux--lithium 
abundance is analyzed for a sample of bona fide subgiant stars, with 
evolutionary status determined from HIPPARCOS trigonometric parallax 
measurements and from the Toulouse--Geneva code. The distribution of 
rotation and CaII emission flux as a function of effective temperature 
shows a discontinuity located rather around the same spectral type, 
F8IV. Blueward of this spectral type subgiants have a large spread 
of values of 
rotation and CaII flux, whereas stars redward of F8IV show essentially low 
rotation and low CaII flux. The strength of these declines depends clearly 
on stellar mass. The abundance of lithium also shows  a sudden decrease. For
subgiants with mass lower than about 1.2~M$_{\odot}$ the decrease is
 located later than  that in rotation and CaII flux, whereas 
for masses higher than 1.2~M$_{\odot}$ the 
decrease in lithium abundance is located around the spectral type F8IV.
 The discrepancy between the location of the discontinuities of rotation
and CaII emission flux and  $\log~n(Li)$ for stars with masses lower 
than 1.2~M$_{\odot}$ seems to reflect the sensitivity of these 
phenomena to the mass
of the convective envelope. The drop in rotation, which results 
mostly from a magnetic braking, requires an increase in the
mass of the convective envelope less than that
required for the decrease in  $\log~n(Li)$.  The location of the 
discontinuity in 
 $\log~n(Li)$ for stars with masses higher than 1.2~M$_{\odot}$, in 
the same region of the discontinuities in rotation and CaII emission
flux, may also be explained by the behavior of the 
deepening of the convective envelope. The more 
massive the star is, the earlier is the increase of the convective envelope. 
In contrast to the relationship between rotation and CaII flux, which is 
fairly linear, the relationship between lithium abundance and rotation 
shows no clear tendency toward  linear behavior. Similarly, no clear
linear trend is  observed in the relationship between lithium abundance and 
CaII flux. In spite of these facts, subgiants with high lithium content also have high 
rotation and high CaII emission flux.
}
  \authorrunning{do Nascimento et al.}
   \titlerunning{Rotation, chromospheric activity and
       Li abundances in subgiant stars}
\maketitle

\section{Introduction}

The study of the influence of stellar rotation on chromospheric activity and
on the mixing of light elements in evolved stars has undergone some important
advances during the past decade. Several authors have reported a 
rotation-activity relation for evolved stars based on the linear behavior of 
the chromospheric flux with stellar rotation (e.g.: Rutten 1987; Rutten and 
Pylyser 1988; Simon and Drake 1989; Strassmeier et al. 1994; Gunn et al. 1998; 
Pasquini et al. 2000).
For a given spectral type, however, a large spread in the rotation--activity
relation is observed, which suggests that rotation might not be the only relevant 
parameter controlling stellar activity.
Indeed, results from  Pasquini and Brocato (1992) and Pasquini et al. (2000) have shown
that chromospheric activity depends on stellar effective temperature and mass.

A possible connection between 
rotation and abundance of lithium in evolved stars 
has also been reported in the literature (e.g.: De Medeiros et al. 1997; 
do Nascimento et al. 2000; De Medeiros et al. 2000). Subgiant and giant 
stars with enhanced lithium abundance show also enhanced rotation, in spite
of a large spread in the abundances of lithium among the slow rotators. In 
addition, do Nascimento et al. (2000) have pointed to a discontinuity in the
distribution of Li abundances as a function of effective temperature later
than the discontinuity in rotation (e.g.:
De Medeiros and Mayor 1990). Concerning the link between chromospheric
activity and light element abundances, Duncan (1981) and Pasquini et al. (1994) 
have found a clear tendency of solar G--type stars with enhanced CaII surface flux $F(CaII)$ 
to have a higher lithium content. This is consistent with the predictions of standard 
evolutionary models, according to which, activity and abundance of light 
elements should depend on stellar surface temperature, metallicity and age. 
In spite of these important studies showing evidence of a connection in between
abundance of lithium and rotation and in between chromospheric activity 
and rotation, in practice, for evolved stars,
the mechanisms controlling such connections and their dependence 
on different stellar parameters like metallicity, mass and age are not yet 
well established.  In this paper, we analyze in parallel the behavior of 
the chromospheric activity, stellar rotation and lithium abundance 
along the subgiant branch. 
In the present approach, the stars are placed in the HR diagram to
determine  more clearly the location of the discontinuities for these 
three stellar parameters based on a sample of bona fide subgiants.

\begin{figure}[t]
\vspace{.2in}
\centerline{\psfig{figure=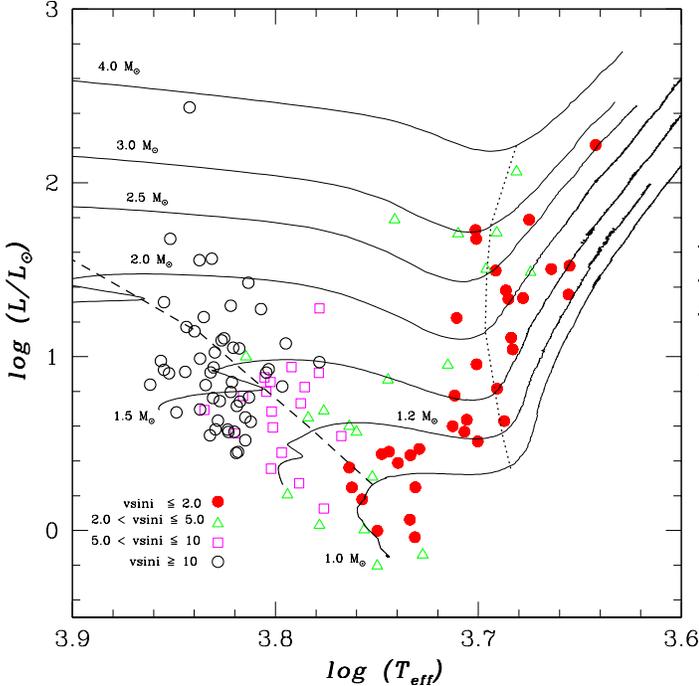,width=3.8truein,height=3.8truein}
}
\caption[]{Distribution of subgiant stars in the HR diagram, with the
rotational behavior as a  function of luminosity and effective
temperature. Luminosities  have been derived from the
HIPPARCOS parallaxes. Evolutionary tracks at [Fe/H]=0 are shown
for stellar masses between 1 and 4~M$_{\odot}$.
The  dashed line indicates the beginning of the  subgiant branch and
the dotted line represents  the  beginning on the red giant branch.
}
\label{FigArtRot}
\end{figure}

\begin{figure}[h]
\vspace{.2in}
\centerline{\psfig{figure=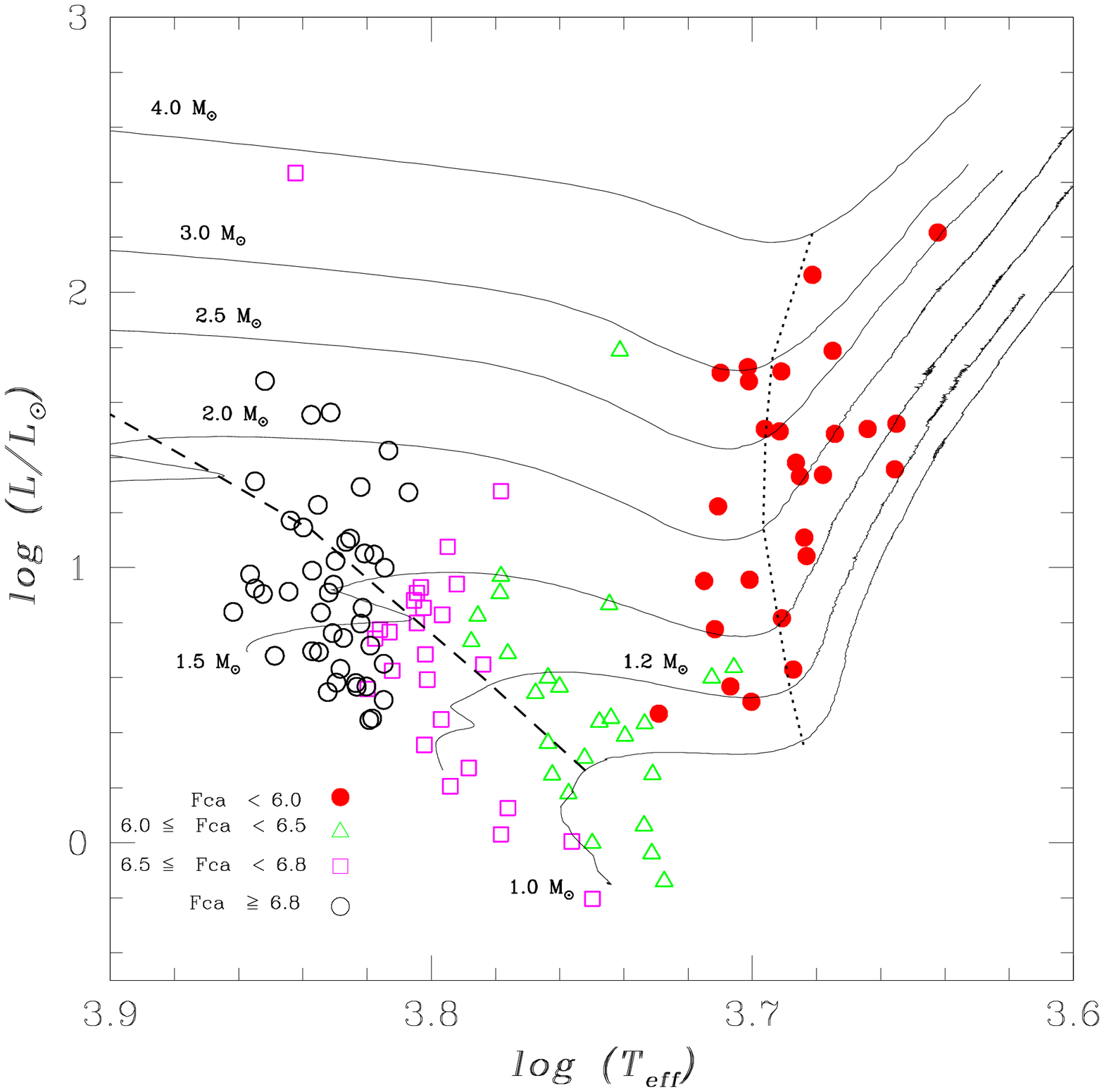,width=3.8truein,height=3.8truein}
}
\caption[]{Distribution of subgiant stars in the HR diagram, with the behavior
of the $F(CaII)$ surface flux as a function of luminosity and effective
temperature.
Luminosities  have been derived from the
HIPPARCOS parallaxes. Evolutionary tracks are defined as in Fig. 1.
}
\label{FigArtCa}
\end{figure}

\begin{figure}[t]
\vspace{.2in}
\centerline{\psfig{figure=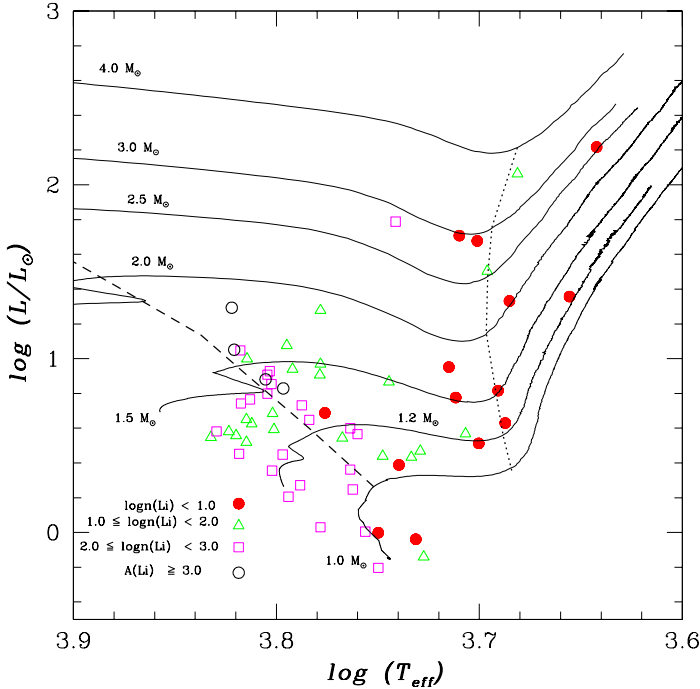,width=3.8truein,height=3.8truein}
}

\caption[]{Distribution of subgiant stars in the HR diagram, with the
behavior of Li abundance as a function of luminosity and effective
temperature.
Luminosities  have been derived from the
HIPPARCOS parallaxes. Evolutionary tracks are defined as in Fig. 1.
}
\label{FigArtLi}
\end{figure}

 \section{Working Sample}

For this study we have selected a large sample of 121 single stars
classified as subgiants in the literature, along the spectral region F, G
and K, with rotational velocity, flux of CaII and $\log~n(Li)$  now
available. The rotational velocities $v\sin i$ were taken from De Medeiros
and Mayor (1999). By using the CORAVEL spectrometer (Baranne et al. 1979)
these authors have determined the projected rotational velocity  $v\sin i$ for
a large sample of subgiant and giant stars with a precision of about 1
km\,s$^{-1}$ for stars with $v\sin i$ lower than about 30 km\,s$^{-1}$.
For higher rotators, the estimations indicate an uncertainty of about
$10\%$. The $F(CaII)$ was determined from the CaII H and K line--core
emission index $S_1$ and $S_2$ listed by Rutten (1987), using the procedure of
conversion from the emission index S1 to flux at the stellar surface $F(CaII)$
given by Rutten (1984). The values of $\log~n(Li)$ were taken from L\`ebre et
al. (1999) and Randich et al. (1999). Readers are referred to these
works for discussion on the observational procedure, data reduction
and error analysis. Stellar luminosities were determined as follows. First, 
the apparent visual magnitudes $m_v$ and trigonometric parallaxes, both 
taken from HIPPARCOS catalogue (ESA 1997), were combined to yield the 
absolute visual magnitude $M_v$.  Bolometric correction $BC$, computed from 
Flower (1996) calibration, was applied giving the bolometric magnitude which
was finally converted into stellar luminosity. The effective temperature 
was computed using Flower (1996) $(B-V)$ versus $T_{eff}$ calibration.
The rotational velocity $v\sin
i$, stellar surface flux $F(CaII)$, abundance of lithium  $\log~n(Li)$
and stellar parameters of the entire sample are presented in 
Table \ref{tabela1}.

\section{Results}

\subsection{The discontinuity in Rotation, CaII emission Flux and Li
abundance}

As a first step, the stellar luminosity and the effective temperature listed in 
Table \ref{tabela1}
were used to construct the HR diagram to better locate the evolutionary stage of 
the stars in the sample. In fact, such a procedure seems important because in preceding 
studies on the link between
rotation and chromospheric activity in subgiant stars, only the spectral type was used as 
a criterion for identifying  the stars. Evolutionary tracks were computed from the 
Toulouse--Geneva code 
for stellar masses
between 1 and 4 M$_{\odot}$, for metallicity consistent with solar--type
subgiant stars (see do Nascimento et al 2000 for a more detailed description).
Here, in particular, we use the evolutionary tracks computed
with  solar metallicity because most of the stars in the present 
sample have $[Fe/H]\sim 0$. The HR diagram with the  evolutionary
tracks is displayed in Figs. \ref{FigArtRot}, \ref{FigArtCa} and \ref{FigArtLi}, which in addition show the
behavior of the rotational velocity $v\sin i$, surface flux CaII  and 
$\log~n(Li)$  abundance respectively. In these diagrams the 
dashed  line indicates
the evolutionary region where the subgiant branch starts, corresponding to
hydrogen exhaustion in stellar central regions, whereas  the  dotted line
represents the beginning of the ascent of the red giant branch. One observes,
clearly, that most of the stars in the present sample are effectively
subgiants. Nevertheless a small number of stars located in particular on the
cool side of the diagrams are rather stars evolving along the red giant
branch. In this context, for the purpose of the present analysis, these 
deviating stars will not be considered as subgiants, in spite of the spectral
 types assigned in the literature. 

\begin{figure}[t]
\vspace{.2in}
\centerline{\psfig{figure=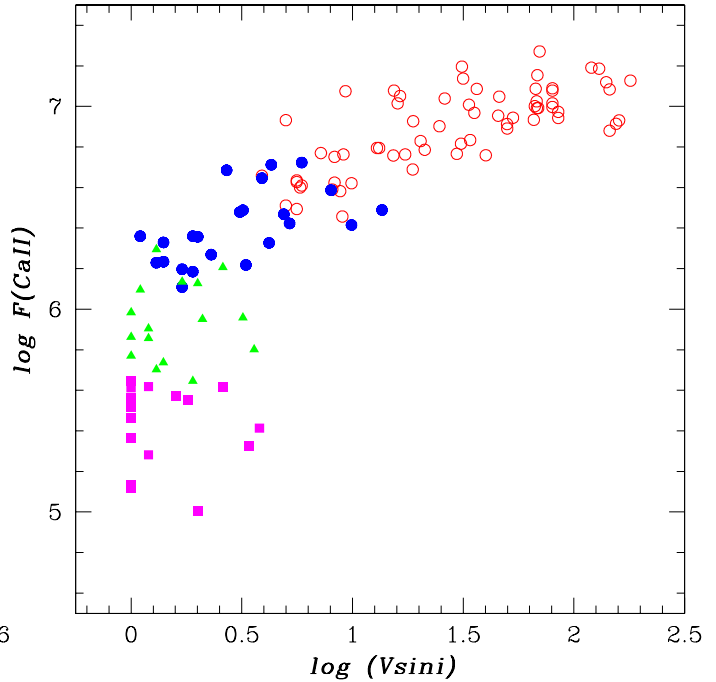,width=3.8truein,height=3.8truein}
}
\caption[]{$log~F(CaII)$ versus $log (vsini)$ for the program stars.
Open circles denote  stars  with $(B-V) \leq 0.55$, filled circles  those with
$ 0.55 < (B-V) \leq 0.75$, triangles  stars  with  $0.75 < (B-V) \leq 0.95$
and squares stars with $(B-V) > 0.95$.  }
\label{FcaVsini}
\end{figure}

\begin{figure}[t]
\vspace{.2in}
\centerline{\psfig{figure=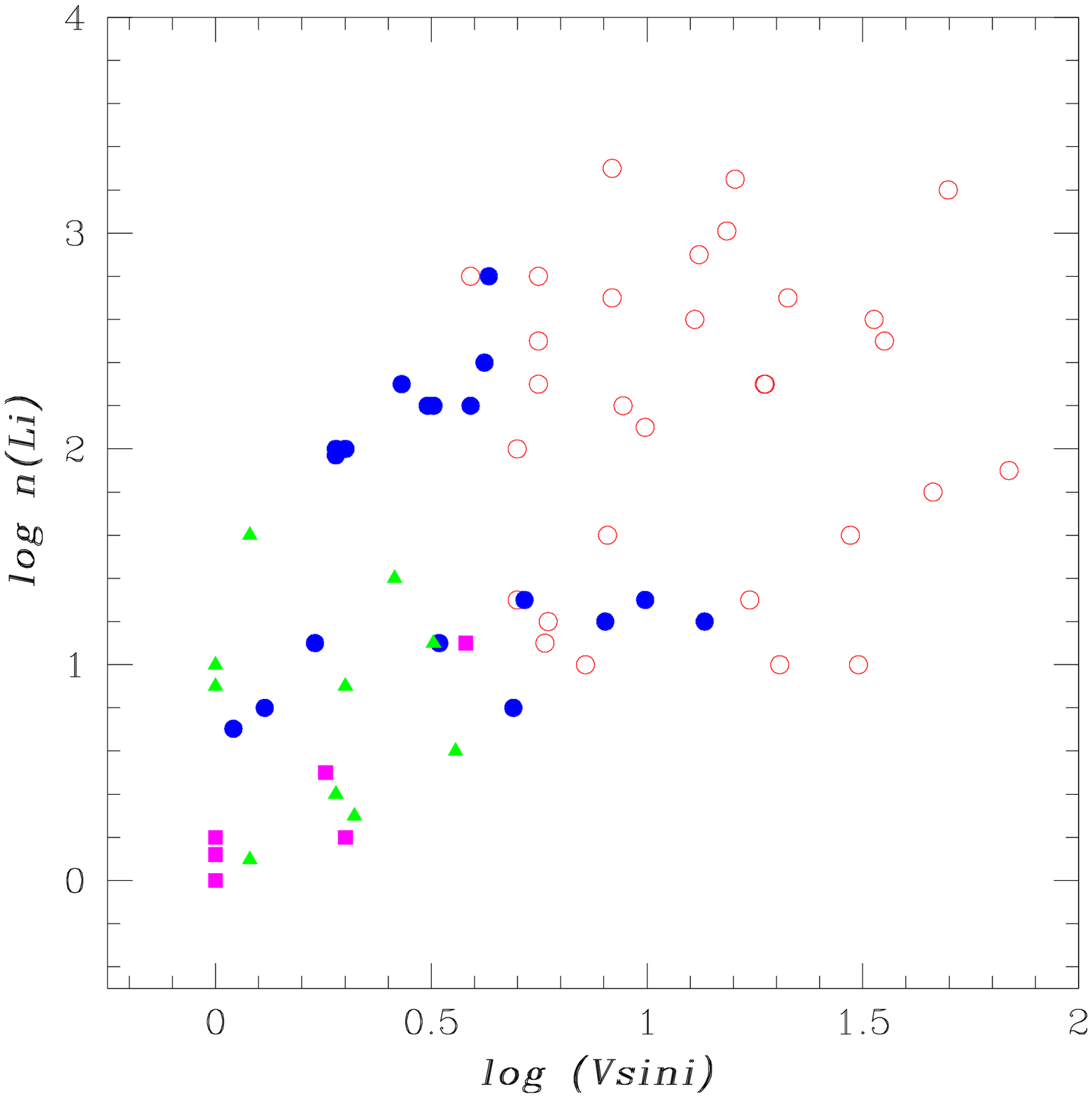,width=3.8truein,height=3.8truein}
}
\caption[]{$\log~n(Li)$ versus $log (vsini)$ for the  program stars.
 Symbols are defined as in Fig. 4 }
\label{ctfig5}
\end{figure}

\begin{figure}[t]
\vspace{.2in}
\centerline{\psfig{figure=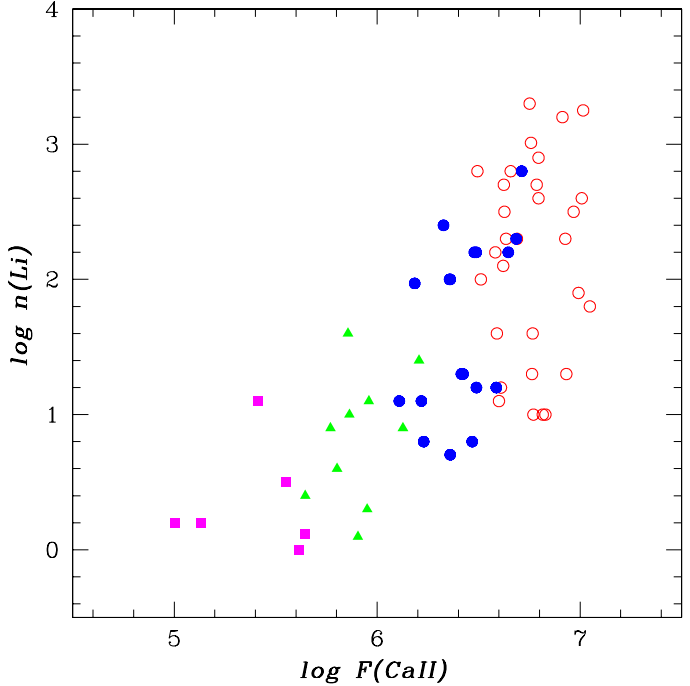,width=3.8truein,height=3.8truein}
}
\caption[]{$\log~n(Li)$ versus $log~F(CaII)$ for the  program stars.
 Symbols are defined as in Fig. 4 }
\label{ctfig6}
\end{figure}

Figure \ref{FigArtRot} shows the well
established rotational discontinuity around the spectral type F8IV 
(e.g.: De Medeiros and Mayor 1990), 
corresponding to $(B-V) \approx 0.55~( \log T_{eff} \sim 3.78$ ). As shown by
these authors, single subgiants blueward of this spectral type show a wide range
of rotational velocities from a few km\,s$^{-1}$ to about one hundred times the
solar rotation, whereas subgiants redward of F8IV are essentially slow rotators,
except for the synchronized binary systems. Fig. \ref{FigArtRot} shows clearly that 
single subgiants redward of the discontinuity with high $v\sin i$ are
unusual. The root cause for such a discontinuity seems to be a strong
magnetic braking associated with the rapid increase of the moment of inertia,
due to evolutionary expansion, once the star evolves along the late F
spectral region (e.g. Gray and Nagar 1985; De Medeiros and Mayor 1990).  

Figure \ref{FigArtCa} shows clear evidence of a discontinuity
in the surface flux $F(CaII)$ paralleling the one
observed in rotational velocity. In fact, such a
sudden decrease in CaII flux of subgiants also parallels that in CIV emission 
flux found by Simon and Drake (1989).  Stars with typical subgiant masses
showing the highest CaII flux are located blueward of this discontinuity.
Such a drop in the surface chromospheric flux is interpreted by Simon and
Drake (1989) as the result of the drop in rotation near the spectral type 
G0IV. According to these authors, there is a development of a dynamo in 
late F stars, which induces a strong magnetic braking in a preexisting wind
that acts on the outermost layers of the stellar surface. As a consequence 
the stellar surface will spin down.

Figure \ref{FigArtLi} shows the behavior of the lithium abundance, 
with a sudden decrease in  $\log~n(Li)$ for subgiant stars with mass
lower than about 1.2 M$_{\odot}$, located a somewhat later than the 
discontinuity in rotation and in surface $F(CaII)$. Evidence 
for this decrease in $\log~n(Li)$ was first pointed out by do
Nascimento et al. (2000). According to these 
authors, such a drop in  $\log~n(Li)$ abundances of subgiants seems 
to result from the rapid increase of the convective envelope  at the late F 
evolutionary stage.  Due to the convective mixing process,  Li--rich 
surface material is diluted
towards the stellar interior. For higher masses, the drop in  
$\log~n(Li)$ shows a tendency to parallel the discontinuities in 
$v\sin i$ and $F(CaII)$, near F8IV, corresponding to 
$(B-V) \approx 0.55~(\log T_{eff} \sim 3.78)$. 

An additional trend is present in Figs. 1 and 2, which show that the fastest rotators
and those subgiants with the highest CaII emission flux, namely the stars blueward of 
F8IV, are mostly stars with mass higher than about 1.2 M$_{\odot}$. Subgiants with mass 
lower than about 1.2 M$_{\odot}$ show moderate to low rotation as well as
 moderate to low surface  $F(CaII)$. In the region blueward of F8IV, 
the abundances of lithium show a more complex behavior for stars with
masses between 1.2 and 1.5  M$_{\odot}$. Fig. 3 shows a number of stars 
in this mass interval with low to moderate  $\log~n(Li)$. Such a fact appears to
reflect the so-called dip region observed by Boesgaard \& Tripicco (1986).

\subsection{The relation Rotation -- $F(CaII)$ -- $\log~n(Li)$}

As a second step of this study we have analyzed the direct relationship
between rotation, $F(CaII)$ and $\log~n(Li)$ for the stars of the sample. Figure 
\ref{FcaVsini} shows the surface  $F(CaII)$ versus the rotational velocity
$v\sin i$, where stars  are separated by intervals of $(B-V)$. Stars earlier
than the rotational discontinuity, typically those with $(B-V) \leq~0.55$,
are represented by open circles,  solid circles stand for stars with
$0.55 < (B-V) \leq 0.75$, triangles stand for stars  with $0.75 < (B-V)$
$\leq  0.95$ and squares represent stars with $(B-V) > 0.95$. The well
established correlation between rotation and chromospheric emission flux 
(e.g. Simon  and Drake 1989), here represented by the  surface  $F(CaII)$, 
is clearly confirmed for the present sample of bona fide subgiants.

Figure 5 presents the behavior of  $\log~n(Li)$ as a function of the rotational
velocity $v\sin i$, confirming the trend of a fair connection in between
abundance of Li and $v\sin i$ in subgiant stars already observed by other 
authors (e.g. De Medeiros et al. 1997).  

Finally, Fig. 6 shows the surface  $F(CaII)$ as a function of  $\log~n(Li)$.
In spite of  more a limited number of stars than in Figs. 4 and 5, we
observe a trend for a connection between $F(CaII)$  and $\log~n(Li)$ following 
rather the behavior observed 
in the $v\sin i$ versus  $\log~n(Li)$ relation.

\begin{figure}[t]
\vspace{.2in}
\centerline{\psfig{figure=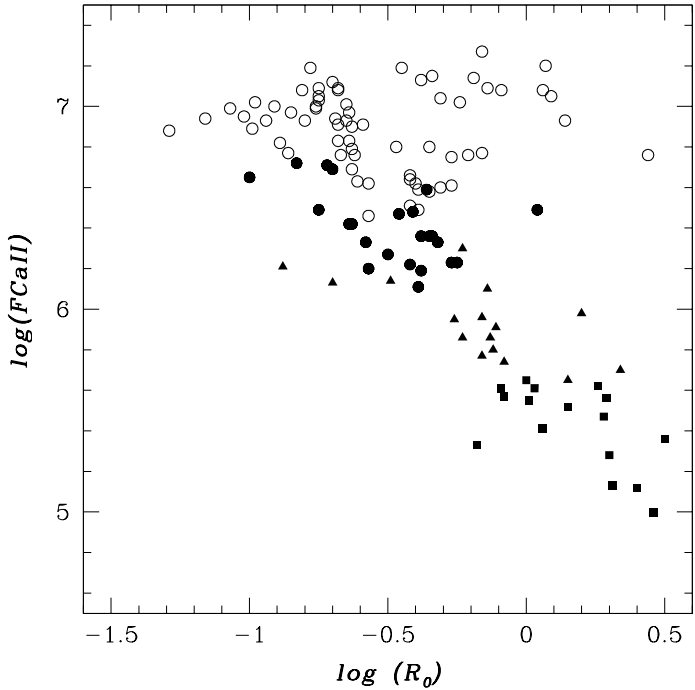,width=3.8truein,height=3.8truein}
}
\caption[]{The $F(CaII)$ versus the Rossby number $R_{\rm 0}$. The 
Symbols are defined as in Fig. 4. }
\label{ro_fcaii}
\end{figure}

\subsection{The connection $F(CaII)$ emission flux--Rossby number}

A close examination of the rotation versus $F(CaII)$ relation presented in Fig. 4 
shows that the amount by which it deviates from a linear correlation depends on the 
(B--V) color interval. A similar color dependence was observed by Noyes et al. (1984), 
who removed such an effect by introducing the dimensionless Rossby 
number $R_{\rm 0}=P_{\rm rot}/\tau_{\rm conv}$, as a 
mesure of the rotational velocity. This dependence was also noted by Simon and Drake (1989) 
for subgiant stars, by analysing the relation $F(CIV)$ versus rotation. These results 
confirm that rotation is not the only parameter expected to influence stellar chromospheric 
activity; another is the stellar mass, or equivalently, the position of the star in 
the HR diagram, which dictates the properties of the stellar convective zone. The deepening 
of the convective zone, or its convective turnover time is, in particular, expected to play 
a relevant role in the dynamo generation. The Rossby number, in fact, determines the extent 
to which rotation can induce both helicity and differential rotation required for 
dynamo activity in the convective zone. 
To analyse the connection $F(CaII)$ emission flux--Rossby number, we have computed $R_{\rm 0}$
for all the stars of the present sample. The convective turnover time $\tau_{\rm conv}$ was estimated from 
the iterated function in (B--V) given by Noyes et al. (1984), whereas the rotation period was 
estimated indirectly from the $v\sin i$ given in Table 1. A statistical correction of $\pi/4$  was 
taken in consideration, to compensate for $\sin i$ effects. The stellar radii were estimated 
following the standard expression as a function of effective temperature 
and luminosity. 
Figure 7 presents the behavior of $F(CaII)$ as a function of the Rossby number $R_{\rm 0}$, with two 
clear different features. For stars with $(B-V) > 0.55$ the correlation of chromospheric 
activity, given by $F(CaII)$, with $R_{\rm 0}$ is significantly better than with rotational 
velocity, whereas stars with $(B-V) \leq~0.55$ show $F(CaII)$ rather uniformly high and 
independent of the $R_{\rm 0}$. A similar result was found by Simon and Drake (1989), by 
analysing the $F(CIV)$ versus $R_{\rm 0}$ relation.

\subsection{The behavior of $\log~n(Li)$ as a function of the deepening of the 
                   convective envelope}
                   
The level of dilution of lithium depends strongly on the level of convection. In this 
context it sounds interesting to analyse the behavior of lithium abundance as a function 
of the deepening of the convective zone for the present sample of stars. For this 
purpose we have estimated the mass of each star $M_{*}$ from the HR diagram presented in 
Sec. 3.1 and then estimated the mass of the convective zone $M_{CZ} $ from an iterated 
function  $M_{CZ} $ ($M_{*}$, Teff) constructed on the basis of the study by do Nascimento et al. 
(2000) on the deepening (in mass) of the convective envelope of evolved 
stars. These authors present the behavior of  $M_{CZ} $ as a function of Teff for stars with 
masses between 1.0 and 4.0 M$_{\odot}$. Figure 8 shows the behavior of $\log~n(Li)$ in the 
 $M_{CZ} / M_{*}$ versus Teff diagram. It is clear that most of the stars with high lithium 
content present an undeveloped convective envelope, whereas stars with low $\log~n(Li)$ 
have a developed convective envelope.

\begin{figure}[t]
\vspace{.2in}
\centerline{\psfig{figure=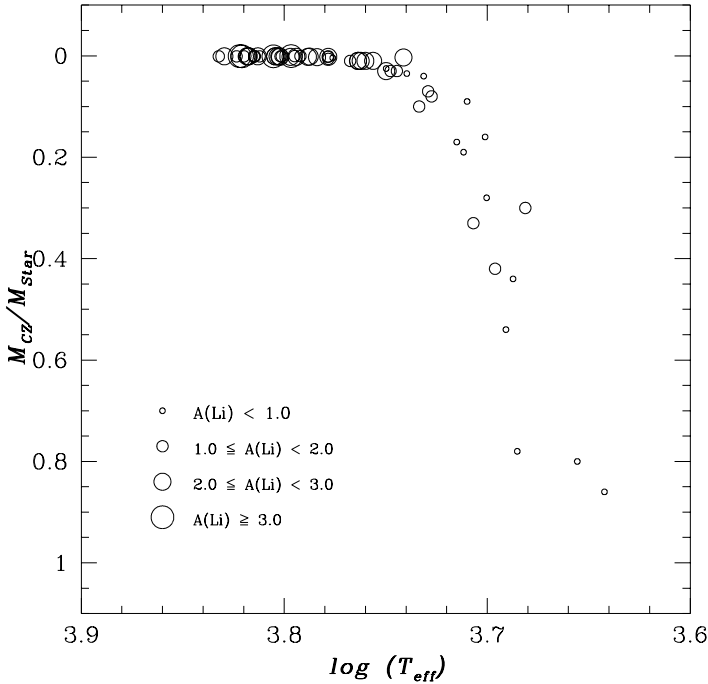,width=3.8truein,height=3.8truein}
}
\caption[]{The deepening (in mass) of the convective envelope as a function 
 of the effective temperature  for the stars in the present sample. 
 The symbol size is proportional to the Li abundances quoted in. }
\label{mzcLi}
\end{figure}

\section{Discussion}

At this point we can inquire about the root cause of the apparent 
discrepancy in the location of the discontinuities in $v\sin i$ and 
$F(CaII)$ and that for  $\log~n(Li)$. In fact, should one expect, from 
the evolutionary point of view, that the discontinuity in  $\log~n(Li)$ 
follows the one in $v\sin i$ as well as that in $F(CaII)$? First of all, 
let us recall that in the specific case of late--type evolved stars, 
chromospheric activity reflects the presence of magnetic fields which are relevant for the 
heating of the chromosphere as well as for mass and angular momentum losses. The 
intensity and spatial distribution of magnetic fields are very probably determined by a 
dynamo process, whose mode of operation and efficiency depends on the interplay between 
stellar rotation and subphotospheric convective motions. In this context one should 
expect  a direct link between the discontinuities in $v\sin i$ and $F(CaII)$, with a drop 
at the same spectral region, if chromospheric activity is directly controlled by 
rotation. {As shown by Fig. 7, this is true, in particular, for stars located redward of the 
spectral region of the discontinuity.}

The question now turns to the apparent discrepancy in the location of the 
discontinuity in  $\log~n(Li)$ in relation to the location of the discontinuities in $v\sin i$ and 
$F(CaII)$ for subgiant stars with masses lower than about 1.2 M$_{\odot}$. 
This discrepancy can be understood as a result of the sensitivity  of these phenomena to the mass of
the convective envelope. In the case of the rotational discontinuity, a small increase in the mass
of the convective envelope is enough to turn the dynamo on. This same dynamo will be responsible
to the magnetic braking causing a drop in the rotation rate and the consequent shutdown of the dynamo
itself. Later, the convective envelope will continue to deep reaching a region previously devoid of 
Li. At this point, the Li brought from the surface layers is diluted and its abundance drops. 
This fact explains clearly the discrepancy between
the location of the  discontinuity in  $\log~n(Li)$ in relation to the one for $vsin i$ and $F(CaII)$, 
as observed from Figs. 1 to 3.  The fact that a magnetic braking might operate with very small changes in the 
mass of the convective envelope is further reinforced by the location of the discontinuity in the 
$F(CaII)$ flux at the late F spectral region. Previous studies (e.g.: do Nascimento et al. 
2000) show that the development of the convective envelope towards the stellar interior 
starts at this spectral region, reaching a maximum within the middle to late G spectral 
region. In short, the drop in $v\sin i$ and 
$F(CaII)$ is earlier than that in  $\log~n(Li)$ because, in contrast to the former, this latter requires 
a large increase in the mass of the convective envelope. Figure 8 shows that Li dilution 
increases abruptly with the deepening of the convective envelope. In fact, the observed discontinuity in  
$\log~n(Li)$ seems to be controlled directly by the increasing of the deepening of the 
convective envelope.

The observed trend for a same location, 
of the  discontinuities in $v\sin i$ 
and  $\log~n(Li)$ for stars with masses larger than about 1.2 M$_{\odot}$ may 
also be explained by following the behavior of the deepening of the convective envelope. 
As shown by do Nascimento et al. (2000, see their Fig. 4), the changes in the mass 
of the convective envelope at a given effective temperature in the range from  
$\log T_{eff}\sim3.75$ to $\log T_{eff}\sim3.68$, are more important for stars with masses in the 
increasing sequence of masses from 1.0 M$_{\odot}$ to 2.5 M$_{\odot}$. The more massive the 
star is, in this range of masses, the earlier is the increasing of the convective envelope. 
In this context, a sudden decrease in  $\log~n(Li)$ of stars with masses larger than about 
1.2 M$_{\odot}$, paralleling the rotational discontinuity, should be expected. 

The relationship between $v\sin i$ and surface  $F(CaII)$, as 
presented in Fig. 4, confirms the results found by other authors for subgiant stars 
(e.g.: Strassmeier et al. 1994) and for other luminosity classes (Strassmeier et al. 1994; 
Pasquini et al. 2000). 
In addition, one observes a trend of increasing scattering in 
the $v\sin i$ versus 
$F(CaII)$ relation, confirming previous claims that rotation might not be the only relevant parameter controlling 
chromospheric activity. In this context, Pasquini et al. (2000) have found for giant 
stars a clear dependence of $F(CaII)$  flux with a high power of stellar effective 
temperature, whereas Strassmeier et al. (1994) have found that the CaII  flux from the cooler evolved stars 
depends more strongly  upon rotation than the CaII flux from the hotter evolved stars. 
The behavior of $F(CaII)$ as a function of the Rossby number $R_{\rm 0}$, presented in Fig. 
7, shows two clear 
trends: For stars with (B--V) larger than about 0.55 the $F(CaII)$ tends towards a linear correlation 
with $R_{\rm 0}$; stars with (B--V) lower than about 0.55 show $F(CaII)$ rather uniformly high 
and independent of $R_{\rm 0}$, pointing for a component of chromospheric activity independent of 
rotation. Different authors (e.g.: Wolff et al. 1986) suggest that the chromospheres of 
early F stars may be heated by the shock dissipation of sound waves, rather than by the dynamo 
process that control the chromospheric activity in G-- and K--type stars.

The dependence of lithium abundance upon rotation observed in Fig. 5 exists in the sense
that the fastest rotators also have the highest lithium content. Nevertheless, there is no 
clear linear relation between these two parameters. Fig. 5 also shows a large spread 
in the Li content at a given $v\sin i$ value, covering at least 2 magnitudes in  $\log~n(Li)$. Such 
a spread shows a clear tendency to increase with rotation and effective temperature. For $v\sin i$ 
lower than about 10 km\,s$^{-1}$, in particular, the  $\log~n(Li)$ values range from 
about 0.0 to about 3.0. Such a spread was also observed by De Medeiros et al. (1997) and
do Nascimento et al. (2000). Finally, the behavior of  $\log~n(Li)$ as a function of 
CaII emission flux presented in Fig. 6 seems to follow roughly  the same trend observed for the 
relation $v\sin i$ versus  $\log~n(Li)$. Subgiants with high lithium content also  show high $F(CaII)$, but
there is  no clear linear relation between these two parameters.

\section{Summary and conclusions}

In the search for a better understanding of the influence of stellar rotation on 
chromospheric activity and lithium dilution, we have analyzed the relationship
rotation--CaII emission flux--Li abundance along the subgiant branch, on the 
basis of a sample of bona fide subgiants, reclassified from HIPPARCOS data.
The evolutionary status of all the stars was determined from trigonometric 
parallax  taken from this  data base and evolutionary tracks computed from the 
Geneva--Toulouse code. 
The distributions of the rotational velocity and of the CaII emission flux
show similar behavior. For both 
parameters we observe a sudden decrease around the spectral type F8IV, 
confirming previous studies. Nevertheless, the extent of these discontinuities 
depends on the stellar mass. Stars with masses around 1.5~M$_{\odot}$ show a more 
important decrease in rotation and CaII emission flux, 
than stars with masses lower than about 1.2~M$_{\odot}$. Clearly, stars blueward of
F8IV, with masses higher than 1.2~M$_{\odot}$, rotate faster and are more active than 
those with masses lower than about 1.2~M$_{\odot}$. The distribution of Li abundance 
versus effective temperature, in spite of a sudden decrease  in the late--F region 
shows a trend for a more complex behavior. First, stars with masses lower than 
about 1.2~M$_{\odot}$ show a discontinuity in  $\log~n(Li)$  somewhat later than the 
discontinuities in rotation and CaII emission flux, whereas stars with higher 
masses present a decline in  $\log~n(Li)$ rather around the spectral type F8IV. In 
addition, a group of stars blueward of F8IV with masses between 1.2 and 1.5~M$_{\odot}$ 
shows moderate to low  $\log~n(Li)$, which seems to reflect the effects of the so-called 
Boesgaard--Tipico dip region. The discrepancy in the location of the 
discontinuities of rotation--CaII emission flux and  $\log~n(Li)$ for stars with masses
lower than 1.2~M$_{\odot}$, seems to be the result of the sensitivity of these phenomena 
to the mass of the convective envelope. The drop in rotation, resulting mostly
from a magnetic braking, requires an increase in the mass of the convective 
envelope less than that required for the sudden decrease  in  $\log~n(Li)$, this later resulting 
from the dilution due to the rapid increase of the convective envelope.
The location of the discontinuity in  $\log~n(Li)$ for stars with masses higher than 
1.2~M$_{\odot}$, in the same region of the discontinuities in rotation and CaII emission 
flux, may also be explained by following the behavior of the deepening of the 
convective envelope. The more massive the star is, the earlier is the increase of the 
convective envelope.  The present work confirms  that the dilution of Li depends strongly 
on the deepening of the convective envelope.

The relationship between rotation and CaII emission flux confirms  previous 
results found by other authors. CaII emission flux shows a correlation with rotation. 
Nevertheless, the large spread in the CaII flux--$v\sin i$ relation 
reinforces previous suggestions that rotation might not be the only relevant 
parameter controlling stellar chromospheric activity. In fact, the relation $F(CaII)$ versus Rossby 
number confirms that chromospheric activity of subgiant stars with (B--V) larger than about 0.55 depends 
rather linearly on rotation, whereas for stars with (B--V) lower than about 0.55 activity is rather 
independent of rotation. 
The relationship between $\log~n(Li)$ and rotation shows a behavior 
less clear than that between CaII flux and 
rotation. Of course the present study confirms a dependence of lithium abundance 
upon rotation, in the sense that stars with the high rotation have also high lithium content. 
In spite of this fact, there is no clear linear relationship
between these two parameters, with a spread more important than that observed in 
the $F(CaII)$ --$v\sin i$ relation. The behavior of the relationship between  lithium abundance and
CaII emission flux seems to follow that observed for  $\log~n(Li)$--$v\sin i$. Stars with 
the high activity also show high lithium content. In both cases there is a
remarkable increase in scattering in the  $\log~n(Li)$--$v\sin i$ and  $\log~n(Li)$--CaII flux 
relations with increasing $v\sin i$ and CaII flux, respectively. Such a fact appears 
to indicate that the influence of rotation on stellar activity is greater than 
on lithium dilution. Finally, the present study point to a pressing need for new 
measurements of chromospheric emission flux and lithium abundance for an 
homogeneous and larger sample of bona fide subgiant stars, with a larger range
of metallicities, than that analyzed 
here. With these additional data it will be possible to analyze the influence of 
rotation upon activity and lithium dilution on a more solid basis, taking into
account the stellar age and metallicity.

\begin{table*}[f]
\begin{center}
\caption{The stars of the present workimg sample with their physical
parameters}
\label{tabela1}
\begin{tabular}{clcrcccccc}\hline\hline
  HD    &  ST            &      log(L/Lo) & $T_{eff}$  &     $v\sin i$  & $F(CaII)$ &   \ $\log~n(Li)$  \\ \hline
  
400	&	F8IV	 &	0.45	&	6265	&	5.6	&	6.635	&	2.30$^{a}$	\\
645	&	K0IV	 &	1.33	&	4844	&	1.8	&	5.551	&	0.50$^{a}$	\\
905	&	F0IV	 &	0.68	&	7059	&	31.6	&	7.137	&			\\	
3229	&	F5IV	 &	1.00	&	6524	&	5.0	&	6.932	&	1.30$^{a}$	\\
4744	&	G8IV	 &	1.49	&	4724	&	3.4	&	5.326	&			\\	
4813	&	F7IV-V	 &	0.21	&	6223	&	3.9	&	6.658	&	2.80$^{a}$	\\
5268	&	G5IV	 &	1.68	&	5024	&	1.9	&	5.646	&	0.40$^{a}$	\\
5286	&	K1IV	 &	1.04	&	4821	&	1.6	&	5.573	&			\\	
6301	&	F7IV-V	 &	0.65	&	6528	&	20.3	&	6.829	&	1.00$^{a}$	\\
6680	&	F5IV	 &	0.63	&	6735	&	36.4	&	7.086	&			\\	
8799	&	F5IV	 &	0.85	&	6628	&	65.9	&	6.934	&			\\	
9562	&	G2IV	 &	0.57	&	5755	&	4.2	&	6.327	&	2.40$^{a}$	\\
11151	&	F5IV	 &	0.80	&	6637	&	34.0	&	6.834	&			\\	
12235	&	G2IV	 &	0.54	&	5855	&	5.2	&	6.423	&	1.30$^{a}$	\\
13421	&	G0IV	 & 	0.91	&	6006	&	9.9	&	6.415	&	1.30$^{a}$	\\
13871	&	F6IV-V	 &	0.77	&	6546	&	9.1	&	6.763	&			\\	
16141	&	G5IV	 &	0.31	&	5653	&	2.3	&	6.269	&			\\	
18262	&	F7IV	 &	0.80	&	6375	&	9.9	&	6.621	&	2.10$^{b}$	\\
18404	&	F5IV	 &	0.57	&	6656	&	24.7	&	6.902	&			\\	
20618	&	G8IV	 &	1.22	&	5137	&	1.0	&	5.984	&			\\	
23249	&	K0IV	 &	0.51	&	5015	&	1.0	&	5.770	&	0.90$^{b}$	\\
25621	&	F6IV	 &	0.83	&	6261	&	15.3	&	6.758	&	3.01$^{b}$	\\
26913	&	G5IV	 &     -0.20	&	5621	&	3.9	&	6.646	&	2.20$^{a}$	\\
26923	&	G0IV	 &	0.03	&	6002	&	4.3	&	6.712	&	2.80$^{a}$	\\
29859	&	F7IV-V	 &	0.83	&	6103	&	9.0	&	6.457	&			\\	
30912	&	F2IV	 &	1.56	&	6877	&    155$^{f}$	&	6.914	&			\\	
33021	&	G1IV	 &	0.36	&	5803	&	2.0	&	6.357	&	2.00$^{a}$	\\
34180   &       F0IV     &      0.74    &       6721    &    80$^{f}$   &       7.015   &                       \\
34411   &       G2IV-V   &      0.25    &       5785    &       1.9     &       6.360   &       2.00$^{a}$      \\
37788   &       F0IV     &      0.92    &       7160    &       31.2    &       7.196   &                       \\
39881   &       G5IV     &      0.18    &       5718    &       1.4     &       6.329   &                       \\
43386   &       F5IV-V   &      0.45    &       6582    &       18.8    &       6.927   &       2.30$^{b}$      \\
53329   &       G8IV     &      1.73    &       5028    &       1.3     &       5.702   &                       \\
57749   &       F3IV     &      2.43    &       6955    &    40$^{f}$   &       6.759   &                       \\
60532   &       F6IV     &      0.94    &       6195    &       8.1     &       6.590   &       1.60$^{a}$      \\
64685   &       F2IV     &      0.70    &       6873    &       67.2    &       7.087   &                       \\
66011   &       G0IV     &      0.97    &       6002    &       13.6    &       6.489   &       1.20$^{a}$      \\
71952   &       K0IV     &      1.11    &       4828    &       1.0     &       5.520   &                       \\
73017   &       G8IV     &      1.50    &       4915    &       1.2     &       5.618   &                       \\
73593   &       G0IV     &      1.38    &       4857    &       1.0     &       5.561   &                       \\
76291   &       K1IV     &      1.50    &       4614    &       1.2     &       5.282   &                       \\
78154   &       F7IV-V   &      0.59    &       6328    &       5.8     &       6.600   &       1.10$^{a}$      \\
81937   &       F0IV     &      1.15    &       6916    &     145$^{f}$ &       7.084   &                       \\
82074   &       G6IV     &      0.95    &       5188    &       2.1     &       5.951   &       0.30$^{a}$      \\
82328   &       F6IV     &      0.88    &       6388    &       8.3     &       6.751   &       3.30$^{a}$      \\
82734   &       K0IV     &      2.06    &       4800    &       3.8     &       5.413   &       1.10$^{a}$      \\
84117   &       F9IV     &      0.27    &       6142    &       5.6     &       6.627   &       2.50$^{b}$      \\
89449   &       F6IV     &      0.63    &       6488    &       17.3    &       6.763   &       1.30$^{a}$      \\
92588   &       K1IV     &      0.57    &       5091    &       1.0     &       5.863   &       1.00$^{a}$      \\
94386   &       K3IV     &      1.36    &       4525    &       1.0     &       5.133   &       0.20$^{a}$      \\
99028   &       F2IV     &      1.05    &       6619    &       16.0    &       7.015   &       3.25$^{b}$      \\
99329   &       F3IV     &      0.91    &       6989    &    130$^{f}$  &       7.186   &                       \\
99491   &       K0IV     &     -0.14    &       5338    &       2.6     &       6.206   &       1.40$^{a}$      \\

\hline
\end{tabular}
\end{center}

Sources: a --  L\`ebre  et al. (1999); b --  De Medeiros et al. (1997); 
c -- Randich  et al. (1999); f --Uesugi and Fukuda (1982);
\end{table*}

\begin{table*}[f]
\begin{center}
{\bf Table 1}.
{Continued. The stars of the present workimg sample with their physical
parameters}
\begin{tabular}{clcrcccccc}\hline\hline
  HD    &  ST            &      log(L/Lo) & $T_{eff}$  &     $v\sin i$  & $F(CaII)$ &   \ $\log~n(Li)$  \\ \hline

104055	&	K2IV	 &	2.22	&	4388	&	2.0	&	5.003	&	0.20$^{a}$	\\
104304	&	K0IV	 &     -0.04	&	5387	&	2.0	&	6.127	&	0.90$^{a}$	\\
105678	&	F6IV	 &	1.08	&	6236	&	29.6	&	6.766	&	1.60$^{a}$	\\
107326	&	F0IV	 &	0.98	&	7185	&    120$^{f}$	&	7.191	&			\\	
110834	&	F6IV	 &	1.27	&	6414	&    145$^{f}$	&	6.880	&			\\	
117361	&	F0IV	 &	1.09	&	6707	&    85$^{f}$	&	6.973	&			\\	
119992	&	F7IV-V	 &	0.36	&	6341	&	8.3	&	6.624	&	2.70$^{a}$	\\
121146	&	K2IV	 &	1.52	&	4520	&	1.0	&	5.116	&			\\	
123255	&	F2IV	 &	1.17	&	6980	&    140$^{f}$	&	7.119	&			\\	
124570	&	F6IV	 &	0.73	&	6130	&	5.6	&	6.494	&	2.80$^{a}$	\\
125111	&	F2IV	 &	0.69	&	6839	&	9.3	&	7.075	&			\\	
125184	&	G5IV	 &	0.39	&	5491	&	1.3	&	6.229	&	0.80$^{a}$	\\
125451	&	F5IV	 &	0.55	&	6796	&	46.0	&	7.048	&	1.80$^{a}$	\\
125538	&	G9IV	 &	1.79	&	4731	&	1.0	&	5.363	&			\\	
126943	&	F1IV	 &	0.99	&	6873	&     80$^{f}$	&	7.078	&			\\	
127243	&	G3IV	 &	1.71	&	5128	&	3.6	&	5.802	&	0.60$^{a}$	\\
127739	&	F2IV	 &	0.94	&	6768	&	68.0	&	6.991	&			\\	
127821	&	F4IV	 &	0.45	&	6596	&	45.5	&	6.954	&			\\	
130945	&	F7IVw	 &	0.93	&	6358	&	18.7	&	6.689	&	2.30$^{b}$	\\
133484	&	F6IV	 &	0.77	&	6502	&	21.2	&	6.786	&	2.70$^{a}$	\\
136064	&	F9IV	 &	0.65	&	6079	&	5.0	&	6.511	&	2.00$^{a}$	\\
143584	&	F0IV	 &	0.84	&	7273	&     70$^{f}$	&	7.271	&			\\	
145148	&	K0IV	 &	0.63	&	4867	&	1.0	&	5.612	&	0.00$^{c}$	\\
150012	&	F5IV	 &	1.05	&	6573	&	35.5	&	6.968	&	2.50$^{a}$	\\
154160	&	G5IV	 &	0.47	&	5360	&	1.2	&	5.856	&	1.60$^{a}$	\\
154417	&	F8.5IV-V &	0.13	&	5972	&	5.9	&	6.723	&			\\	
156697	&	F0-2IV-Vn &	1.56	&	6782	&     160$^{f}$	&	6.931	&			\\	
156846	&	G3IV	 &	0.69	&	5972	&	4.9	&	6.468	&	0.80$^{a}$	\\
157347	&	G5IV	 &	0.00	&	5621	&	1.1	&	6.360	&	0.70$^{c}$	\\
157853	&	F8IV	 &	1.79	&	5511	&	3.2	&	6.488	&	2.20$^{a}$	\\
158170	&	F5IV	 &	1.28	&	6002	&	8.0	&	6.587	&	1.20$^{a}$	\\
161797	&	G5IV	 &	0.43	&	5414	&	1.7	&	6.109	&	1.10$^{a}$	\\
162003	&	F5IV-V	 &	0.74	&	6569	&	12.9	&	6.795	&	2.60$^{a}$	\\
162076	&	G5IV	 &	1.50	&	4967	&	3.2	&	5.959	&	1.10$^{a}$	\\
162917	&	F4IV-V	 &	0.57	&	6610	&     50$^{f}$	&	6.891	&			\\	
164259	&	F2IV	 &	0.76	&	6772	&     80$^{f}$	&	6.997	&			\\	
165438	&	K1IV	 &	0.82	&	4907	&	1.0	&	5.647	&	0.12$^{c}$	\\
173949	&	G7IV	 &	1.71	&	4909	&	2.6	&	5.614	&			\\	
176095	&	F5IV	 &	0.91	&	6375	&	13.2	&	6.795	&	2.90$^{a}$	\\
182572	&	G8IV	 &	0.25	&	5384	&	1.7	&	6.135	&			\\	
182640	&	F0IV	 &	0.90	&	7119	&	68.4	&	7.154	&			\\	
184663	&	F6IV	 &	0.58	&	6660	&       69.0	&	6.992	&	1.90$^{b}$ \\
185124  &       F3IV     &      0.72    &       6592    &     85$^{f}$  &       6.943   &                       \\
188512  &       G8IV     &      0.78    &       5148    &       1.2     &       5.905   &       0.10$^{c}$      \\
190360  &       G6IV+M6V &      0.06    &       5417    &       1.7     &       6.197   &                       \\
190771  &       G5IV     &      0.01    &       5705    &       2.7     &       6.685   &       2.30$^{a}$      \\
191026  &       K0IV     &      0.60    &       5160    &       1.3     &       6.295   &                       \\
191570  &       F5IV     &      0.58    &       6754    &       33.6    &       7.008   &       2.60$^{a}$      \\
192344  &       G4IV     &      0.45    &       5547    &       1.4     &       6.234   &                       \\
195564  &       G2.5IV   &      0.44    &       5593    &       1.9     &       6.185   &       1.97$^{c}$      \\
196755  &       G5IV+K2IV &     0.87    &       5553    &       3.3     &       6.218   &       1.10$^{a}$      \\
197373  &       F6IV     &      0.52    &       6528    &       30.9    &       6.816   &       1.00$^{a}$      \\
197964  &       K1IV     &      1.34    &       4764    &       1.0     &       5.466   &                       \\

\hline
\end{tabular}
\end{center}

Sources: a --  L\`ebre  et al. (1999); b --  De Medeiros et al. (1997); 
c -- Randich  et al. (1999); f --Uesugi and Fukuda (1982);
\end{table*}

\begin{table*}[f]
\begin{center}
{\bf Table 1}.
{Continued. The stars of the present workimg sample with their physical
parameters}
\begin{tabular}{clcrcccccc}\hline\hline
  HD    &  ST            &      log(L/Lo) & $T_{eff}$  &     $v\sin i$  & $F(CaII)$ &   \ $\log~n(Li)$  \\ \hline

198149	&	K0IV	 &	0.96	&	5022	&	1.4	&	5.737	&			\\	
201507	&	F5IV	 &	1.23	&	6844	&	16.4	&	7.051	&			\\	
201636	&	F3IV	 &	0.91	&	6791	&	67.9	&	7.025	&			\\	
202444	&	F1IV	 &	1.02	&	6758	&	26.1	&	7.039	&			\\	
202582	&	G2IV+G2IV &	0.60	&	5803	&	3.1	&	6.479	&	2.20$^{a}$	\\
205852	&	F1IV	 &	1.68	&	7109	&    180$^{f}$	&	7.127	&			\\	
207978	&	F6IV-Vvw &	0.56	&	6605	&	7.2	&	6.770	&	1.00$^{a}$	\\
208703	&	F5IV	 &	0.84	&	6829	&	15.4	&	7.078	&			\\	
210210	&	F1IV	 &	1.31	&	7160	&    80$^{f}$	&	7.089	&			\\	
212487	&	F5IV	 &	0.85	&	6345	&	8.8	&	6.582	&	2.20$^{b}$	\\
216385	&	F7IV	 &	0.68	&	6336	&	5.9	&	6.610	&	1.20$^{a}$	\\
218101	&	G8IV	 &	0.64	&	5078	&	1.1	&	6.096	&			\\	
219291	&	F6IVw	 &	1.43	&	6506	&	53.1	&	6.944	&			\\	
223421	&	F2IV	 &	1.11	&	6688	&	66.6	&	7.001	&			\\	
224617	&	F4IV	 &	1.29	&	6637	&	49.9	&	6.913	&	3.20$^{b}$	\\

\hline
\end{tabular}
\end{center}

Sources: a --  L\`ebre  et al. (1999); b --  De Medeiros et al. (1997); 
c -- Randich et al. (1999); f --Uesugi and Fukuda (1982);
\end{table*}

 \begin{acknowledgements}
 This work has been supported by continuous grants from
 the CNPq Brazilian  Agency.  J.D.N.Jr. acknowledges the
  CNPq grant PROFIX 540461/01-6. Special thanks to the referee, Dr. R. Cayrel for very useful comments, which 
greatly improved the quality of this paper.

  \end{acknowledgements}

\end{document}